\documentclass[twocolumn,final]{revtex4-1}
\usepackage[utf8]{inputenc}
\usepackage{amsmath}
\usepackage{amsfonts}
\usepackage{amssymb}
\usepackage{graphicx}
\usepackage[ngerman,british]{babel}
\usepackage{color}
\let\oldsqrt\sqrt
\def\sqrt{\mathpalette\DHLhksqrt}
\def\DHLhksqrt#1#2{%
\setbox0=\hbox{$#1\oldsqrt{#2\,}$}\dimen0=\ht0
\advance\dimen0-0.2\ht0
\setbox2=\hbox{\vrule height\ht0 depth -\dimen0}%
{\box0\lower0.4pt\box2}}

\begin{document}
\title{Channelization in Porous Media driven by Erosion and Deposition}

\author{R. Jäger} 
\email{jaegerr@ethz.ch} 
\affiliation{  ETH
  Z\"urich, Computational Physics for Engineering Materials, Institute
  for Building Materials, Wolfgang-Pauli-Strasse 27, HIT, CH-8093 Z\"urich
  (Switzerland)}

\author{M. Mendoza} 
\email{mmendoza@ethz.ch} 
\affiliation{  ETH
  Z\"urich, Computational Physics for Engineering Materials, Institute
  for Building Materials, Wolfgang-Pauli-Strasse 27, HIT, CH-8093 Z\"urich
  (Switzerland)}
  
\author{H. J. Herrmann}
\email{hjherrmann@ethz.ch} 
\affiliation{  ETH
  Z\"urich, Computational Physics for Engineering Materials, Institute
  for Building Materials, Wolfgang-Pauli-Strasse 27, HIT, CH-8093 Z\"urich
  (Switzerland)}

\begin{abstract}
We develop and validate a new model to study simultaneous erosion and deposition in three-dimensional porous media. We study the changes of the porous structure induced by the deposition and erosion of matter on the solid surface and find that when both processes are active, channelization in the porous structure always occurs. The channels can be stable or only temporary depending mainly on the driving mechanism. Whereas a fluid driven by a constant pressure drop in general does not form steady channels, imposing a constant flux always produces stable channels within the porous structure. Furthermore we investigate how changes of the local deposition and erosion properties affect the final state of the porous structure, finding that the larger the range of wall shear stress for which there is neither erosion nor deposition, the more steady channels are formed in the structure.

\end{abstract}
\maketitle
\section{Introduction}

Fluid flow through a porous medium can erode and/or deposit material and thereby change the shape of the solid boundaries, which in turn alter the path of the flow. These processes are responsible for shaping a variety of landscapes, some examples are meandering rivers \cite{middleton2005encyclopedia}, coastal erosion \cite{masselink2014introduction} and seepage \cite{cedergren1977seepage}. In industrial applications these processes also play a crucial role and are sometimes desired and sometimes disruptive, e.g. mechanical filters take advantage of deposition whereas in oil wells erosion produces sand that can damage the pumps, requiring expensive counter-measures. While these processes are easily observed in surface flows, it is much more difficult to do so when they occur within. Albeit there is an abundance of applications, the evolution of structures exposed to erosion and deposition is in general not easily predictable and usually requires the use of numerical techniques. 

Erosion is the mechanical wearing of solid material by the shear force exerted by the fluid. The erosion rate is assumed to be proportional to the magnitude of the wall shear force \cite{FLM:57479,Bonelli2006555}. For the erosion and deposition two thresholds for the shear force are introduced. Above a critical threshold the shear stress is high enough to erode matter from the solid surface whereas below a lower threshold the adhesive force of the suspended particles is dominant and deposition occurs. In the range between the thresholds there is locally no net change of mass at the surface. While this is an heuristic approach, it is well-founded, as it is consistent with the Hjulström curve \cite{Grabowski2011101} which describes a gap between the erosion and the sedimentation flow velocity for all grain sizes. To accurately describe the erosion and deposition induced by the wall shear force and adhesive forces within a porous media, the Navier-Stokes equations must be solved and the trajectories of suspended particles have to be calculated. A commonly used approach is to describe the suspended particles as a solute \cite{SALLES19932839, Bouddour1996, yamamoto2009fluid}, whose evolution is given by the convection-diffusion equation. Even though these equations can be written down easily, solving them with realistic boundaries defined by the porous structure cannot be done analytically. Hence we develop a numerical model based on the lattice Boltzmann method (LBM), which is a very powerful tool to resolve fluid behavior in flows through porous media \cite{talon2012assessment}, that allows to describe the erosion induced by wall shear stress and also the deposition of fluid entrained matter.

Using our model we find that when both erosion and deposition are active, channels within the porous structure form. Mahadevan et al. \cite{mahadevan2012flow} and Kudrolli et al. \cite{PhysRevLett.117.028001} already studied channelization in porous media using experiments and two-dimensional effective models for deposition and erosion. In both cases the fluid is modeled using Darcy's law, which corresponds to a scale much larger than the pore size, and therefore, prohibiting the use of local shear stress forces at the solid boundaries to characterize the erosion and deposition processes. In particular, in the model of Mahadevan et al. the erosion depends on the pressure gradient of the fluid and Kudrolli et al. consider erosion and deposition depending on fluid velocity. In our model however, the Navier-Stokes equations are solved to resolve the fluid flow at pore scale and erosion can be considered to depend on the local shear stress exerted by the fluid onto the solid surface. Our goal is to study the microscopic physics of erosion and deposition and the resulting macroscopic changes of the three-dimensional porous structure. Since we want to study the interplay between erosion and deposition we only consider consolidated porous media; and hydraulic pressures that are high enough to erode material, but do not lead to a decompaction or fluidization of the porous medium \cite{PhysRevE.78.051302}.

This paper is organised as follows: in Sec. \ref{sec:theory} we present the set of differential equations needed to understand the erosion and deposition processes in porous media; in Sec.  \ref{sec:modeldescription}, we introduce our model and perform some validations; and Sec. \ref{sec:results} shows the results of our study on erosion and deposition in porous media including the channelization process. Finally, in Sec. \ref{conclusions} we summarize our results and conclusions.

\section{Theory}
\label{sec:theory}

\subsection{Fluid and Particle Dynamics}
There are several mechanisms that are responsible for the mechanical erosion and deposition and have to be considered to get the full description of evolution of the porous structure. First, the fluid flowing through a porous medium is described by the incompressible Navier-Stokes equations, which consist of the continuity equation,
\begin{equation}
\label{eq:cont}
	\frac{\partial\rho}{\partial t} + \nabla \cdot (\rho \vec{u}) = 0 ,
\end{equation} 
and momentum conservation equation
\begin{equation}
\label{eq:mom}
	\frac{\partial \vec{u}}{\partial t} + (\vec{u} \cdot \nabla) \vec{u} - \nu \Delta \vec{u} = -\frac{\nabla p}{\rho}  + \vec{f}, 
\end{equation}
where $\vec{u}$ is the velocity field of the fluid, $\rho$ is the fluid density, $\nu$ the kinematic viscosity, $p$ the hydrostatic pressure, and $\vec{f}$ an acceleration due to external forces. In our approach, the fluid contains suspended particles that are entrained by the fluid and can be deposited onto the solid surface.  The suspended particles that we consider are small, get transported by the fluid, diffuse within it and are neutrally buoyant. Therefore they are described by the convection-diffusion equation, which yields the mass concentration of solid particles in the fluid. This equation reads:
$$
\frac{\partial \mathcal{C}}{\partial t} + \nabla \cdot(\mathcal{C}\vec{u}) = \nabla \cdot(D\nabla \mathcal{C}),
$$
where $D$ is the diffusion coefficient and $\mathcal{C}$ is the particle concentration. Note that the concentration inside the porous medium will also change due to deposition and erosion as we will describe in the following section. Furthermore the concentration is assumed to be low, such that the change in rheology and the momentum exchange between fluid and suspended particles is negligible. By solving the Navier-Stokes and the convection-diffusion equations, we describe the evolution of the fluid and the suspended particles through the porous media.

\subsection{Erosion and Deposition}
\label{sec:erdep}
The wall shear stress (WSS) exerted by the fluid on a solid surface is defined by:
\begin{equation}
	\tau_w \equiv \tau (y= 0) = \mu \frac{\partial u}{\partial y} \bigg|_{y=0} ,
\end{equation} 
where $\mu = \rho\nu$ is the dynamic viscosity, $u$ the fluid velocity parallel to the surface, and $y$ the distance from the surface.
When the WSS overcomes the cohesive forces of the solid, solid matter is eroded and entrained by the fluid. On the other hand, when the adhesive forces, e.g. Van der Waals forces, are stronger than the erosive and other repulsive forces, particles within the fluid are deposited on the solid surface. Through these processes the solid surface and thus the boundary is altered and the flow changes. We consider slow erosion and deposition, where the change of the surface is much slower than the fluid flow velocity. Hence, the boundary can be considered quasi-static (or evolving slowly), i.e. non-moving, and only the magnitude of the deviatoric shear stress and not the hydraulic pressure has to be taken into account for the evolution of the surface. The shear force is calculated from the deviatoric shear tensor given by:

\begin{equation}
	\sigma_{ab} \equiv \mu \bigg( \frac{\partial u_a}{\partial x_b} + \frac{\partial u_b}{\partial x_a} \bigg), 
\end{equation}
multiplied with the surface normal vector $\hat{n}$, yielding the shear force $\vec{\tau} = \hat{n}\cdot \sigma$. The WSS is now the tangential part of the shear force which is obtained with the following expression:
\begin{equation}
\label{eq:tang}
\tau_w = \sqrt{(\hat{n}\cdot\sigma)^2 - ( \hat{n}(\hat{n}\cdot\sigma))^2} ,
\end{equation}
the details for calculating the WSS are shown in section \ref{sec:calcwss}. In the case of erosion, we assume that this WSS works as a wearing force producing sand or silt out of solid matter which is entrained with the flow. The erosion is assumed to be linearly proportional to the wall shear stress, this erosion law is widely used and experimentally verified \cite{Bonelli2006555,FLM:57479}. The erosion is proportional to the stress gap, i.e. the WSS excess above some critical threshold and depends on the density and toughness of the solid matter. This dependency is incorporated in an erosion coefficient $\kappa_{er}$ and a critical shear stress $\tau_{er}$ below which no erosion occurs. The porous medium is considered consolidated and incompressible such that the normal component of the stress tensor can be neglected and other erosive processes, such as fluidization \cite{PhysRevE.78.051302} are not considered. For the deposition law we also assume a threshold of the WSS below which deposition occurs. The deposition of matter is also considered linearly proportional to the WSS below the threshold, hence at the critical deposition shear stress $\tau_{dep}$ no deposition occurs. The deposition is naturally dependent on the concentration $\mathcal{C}$ of solid matter in the fluid. This gives us an erosion and deposition law, which written in differential form, looks as follows:

\begin{equation}
	\dot{m} =
	\begin{cases}
		-\kappa_{er} (\tau_w - \tau_{er}) , \qquad  & \tau_w > \tau_{er} \quad, \\
		+ \frac{\mathcal{C}}{\mathcal{C}_0} \cdot \kappa_{dep} (\tau_{dep}-\tau_w) , \qquad & \tau_w < \tau_{dep} \quad , \\
		0 , \qquad &\tau_{er} \geq \tau_w \geq \tau_{dep} \quad ,
	\end{cases}
\end{equation}

where $\dot{m}$ is the change in mass per unit area and $\mathcal{C}_0$ is the fully saturated concentration. In the following the relative concentration $\mathcal{C}/\mathcal{C}_0$ will be simply referred to as $C$. While the linear relation of erosion to wall shear stress is an empirical law, it is well established and reasonable for a variety of materials such as cohesive soils, clay and materials of similar properties \cite{FLM:57479,Bonelli2006555,bonelli2013erosion}. Furthermore, as we are considering wall shear stresses which are in the vicinity of the critical shear stress, higher order terms would only contribute minimally to erosion. The deposition of particles is usually assumed to be proportional to the concentration $ \propto C$ \cite{SALLES19932839,Bouddour1996,yamamoto2009fluid}, when the solubility is zero, however recent experimental findings \cite{Alem2015} show that deposition decreases with increasing flow speed, thus we chose the simplest form that takes this into account, but converges to $\propto C$ for low flow speeds.
The erosion coefficient $\kappa_{er}$ and the deposition coefficient $\kappa_{dep}$ depend on the properties of the solid matter, e.g. the cohesive forces \cite{Grabowski2011101}.

\section{Numerical Implementation}
\label{sec:modeldescription}

Our numerical model is based on the lattice Boltzmann method (LBM) \cite{wolf2000lattice} with an improved two-relaxation-time (TRT) collision operator \cite{PhysRevE.83.056710,talon2012assessment}. In order to compute accurately solid boundaries and shear stress forces we also implement and analyze different interpolation and extrapolation schemes \cite{Filippova1998219,bouzidi,Mei2000680}. In particular, this avoids the use of the stair-case approximation for curved boundaries. In this work, parameters are written in numerical units, the conversion from numerical to physical units is illustrated in the appendix \ref{app:units}.

\subsection{Lattice Boltzmann Method}
\label{sec:LBM}
To resolve the fluid flow at pore scale we use the D3Q19 lattice Boltzmann method (LBM) \cite{wolf2000lattice}, which means the model is based on a three-dimensional lattice with nineteen discrete velocity vectors $\{\vec{c}_0,...,\vec{c}_{18}\}$. In contrast to other computational fluid methods that solve the Navier-Stokes equations directly, the LBM solves the underlying discrete Boltzmann equation \cite{FLM:409537}: 
\begin{equation}
\label{eq:genlbm}
		f_i(\vec{x}+\vec{c}_i, t+1 ) - f_i(\vec{x},t) = \Omega(\vec{f}),
\end{equation}
where $f_i$ are distribution functions associated with the discrete velocity vectors $\vec{c}_i$, and a generic collision operator $\Omega$. Considering the Bhatnagar-Gross-Krook (BGK) approximation for the collision operator the lattice Boltzmann equation gets the following form \cite{FLM:409537}:
\begin{equation}
\label{eq:lbm}
f_i(\vec{x}+\vec{c}_i, t+1 ) - f_i(\vec{x},t) = -\frac{1}{\mathcal{T}}\left[f_i - f_i^{eq} \right],
\end{equation}
where $\mathcal{T}$ is a relaxation time, that is related with the kinematic viscosity of the fluid by $\nu = (\mathcal{T}-1/2)c_s^2$. The equilibrium distribution function is given by \cite{wolf2000lattice}:
\begin{equation}
\label{eq:feq}
	f_i^{eq} = \omega_i \rho \bigg \lbrace 1+ 3 \frac{(\vec{c}_i \cdot \vec{u} )}{c_s^2} + \frac{9}{2}\frac{(\vec{c}_i \cdot \vec{u})^2} {c_s^4} -\frac{3}{2} \frac{\vec{u}^2}{c_s^2} \bigg\rbrace ,
\end{equation}
where $\rho$ and $\vec{u}$ are the local density and velocity of the fluid, $\omega_i$ are weights (associated with the discrete velocity vectors) and the speed of sound is given by $c_s^{2}=1/3$. The values of the weights $\omega_i$ and discrete velocities $\vec{c}_i$ can be found in Ref. \cite{wolf2000lattice}.

The fluid density is computed from the zeroth moment of the distribution function, i.e. $\rho = \sum_i f_i$, and the fluid velocity from the first moment, $\vec{u} = \sum_i f_i \vec{c}_i /\rho$. Using the Chapman-Enskog expansion it can be shown that Eq.~\eqref{eq:lbm} with Eq.~\eqref{eq:feq} converges to the Navier-Stokes equations Eqs.~\eqref{eq:cont} and \eqref{eq:mom} in the continuum limit \cite{wolf2000lattice}.

\subsection{Improved Collision Operator}
\label{sec:trt}
The algorithm for the generic lattice Boltzmann equation \eqref{eq:genlbm} can be split into collision part
$$
	f^*_i(\vec{x},t) = f_i(\vec{x},t) + \Omega(\vec{f}),
$$
and streaming part
$$
f_i(\vec{x}+\vec{c}_i, t+1 ) = f^*_i(\vec{x},t).
$$

For the BGK collision operator (see Eq.~\eqref{eq:lbm}) the discretization error is viscosity dependent \cite{Pan2006898}, however this problem can be mitigated in two different ways; one is to simply increase the resolution which will decrease the error; and the other is to use a multi-relaxation-time (MRT) collision operator \cite{Pan2006898}. The two-relaxation-time (TRT) collision operator is by far the most efficient and its accuracy for porous media is as good as any other MRT method \cite{PhysRevE.83.056710,talon2012assessment}. The TRT method splits the distribution functions into symmetric and asymmetric parts $f^\pm_i = (f_i \pm f_{\bar{i}}) /2$, where $\bar{i}$ is the index of the velocity vector pointing in the opposite direction of $i$, i.e. $\vec{c}_{\bar{i}} = - \vec{c}_i$. The projected equilibrium distribution functions read $f^{eq,\pm}_i = (f^{eq}_i \pm f^{eq}_{\bar{i}}) /2$. 

For the TRT method the full collision operator $\Omega(\vec{f})$ reads:

\begin{equation}
\label{eq:trt}
\begin{split}
f^*_i = [f_i - \mathcal{T}^{-1} (f^+_i - f^{eq,+}_i) - \mathcal{T}'^{-1} (f^-_i-f^{eq,-}_i)], \\
\end{split}
\end{equation}

where the second relaxation time $\mathcal{T}'$ is a free parameter relaxing the asymmetric part of the distribution functions. To minimize the dependency of the viscosity the parameter $\Lambda \equiv (\mathcal{T}-1/2)(\mathcal{T}'-1/2)$ is chosen to be constant ($\Lambda=0.1$) and therefore the second relaxation time is determined as $\mathcal{T}'=1/2 + \Lambda/(\mathcal{T}-1/2)$ \cite{porescalelbm}. The collision for the distribution functions $f_i$ and $f_{\bar{i}}$ can be calculated simultaneously via Eq.~\eqref{eq:trt}, which allows the implementation of the TRT to be almost as fast as the one of the single-relaxation-time (BGK) approach.

\subsection{Solving the Convection-Diffusion Equation}
To model the convection-diffusion with the lattice Boltzmann method a second set of distribution functions $g_i$ can be used to describe the mass concentration of the solute \cite{Yamamoto2584} within the solvent: 
\begin{equation}
	g_i(\vec{x}+\vec{c}_i, t+1 ) - g_i(\vec{x},t) = -\frac{1}{\mathcal{T}_s}\left[g_i - g_i^{eq} \right] \label{eq:diff}
\end{equation}
Here, the equilibrium distribution is similar to the one for the fluid (see Eq. \eqref{eq:feq}) with the concentration instead of the density:
$$
	g_i^{eq} = \omega_i C \bigg \lbrace 1+ 3 \frac{(\vec{c}_i \cdot \vec{u} )}{c_s^2} + \frac{9}{2}\frac{(\vec{c}_i \cdot \vec{u})^2} {c_s^4} -\frac{3}{2} \frac{\vec{u}^2}{c_s^2} \bigg \rbrace .
$$
The diffusion coefficient is related to the relaxation time $\mathcal{T}_s$, $D = (\mathcal{T}_s - 1/2)c_s^2$. For the TRT method the evolution of the solution is calculated analogously to Eq. \ref{eq:trt} with the symmetric and asymmetric splitting $g_i^\pm$ and the second relaxation parameter $\mathcal{T}'_s = 1/2 + \Lambda/(\mathcal{T}_s - 1/2)$. Analogously to the calculation of the fluid density, the zeroth moment of the concentration distribution function $g_i$ is the concentration of the solute, $C = \sum_{i}g_i$, and via a Chapman-Enskog expansion, one can show that the lattice Boltzmann equation \eqref{eq:diff} converges to the convection-diffusion equation. Note that the positivity of the kinematic viscosity and diffusivity requires $\mathcal{T},\mathcal{T}_s > 1/2$, and the model becomes unstable for $\mathcal{T} \to 1/2$. For the simulations shown in this paper we only used the relaxation times 1 and 0.6, however we found our model stable for values down to 0.503.

\subsection{Solid mass field to identify fluid and solid domains}
\label{sec:mass}
To distinguish fluid and solid domains we introduce a new variable, a scalar mass field that is defined on the lattice, for fluid nodes the mass is zero, for solid nodes the mass is one and for interface cells the mass is between zero and one:
$$
	m(\vec{x}) =
	\begin{cases}
		1 , \quad & \text{solid node} \\
		0 , \quad & \text{fluid node} \\
		(0, 1] . \quad & \text{interface node}
	\end{cases}
$$
If the mass is one, the cell is still an interface node if there is a neighboring fluid node. Interface nodes have both solid and fluid matter and therefore also non-zero distribution functions $f_i > 0$. When an interface node becomes a solid node the distributions are set to zero $f_i = 0 $. When a solid node becomes interface the distribution functions are linearly extrapolated so as not to introduce an artificial pressure or velocity gradient as was done by Yin et al. \cite{yin}:
$$
	f_i(\vec{x}) = \frac{\sum_i \omega_i\left[ 2 f_i(\vec{x} + \vec{c}_i) - f_i(\vec{x}+2\vec{c}_i) \right]}
	{\sum_i \omega_i},
$$
where the sums run over all neighbors $(\vec{x}+\vec{c}_i)$ that are fluid nodes. 

When the solute is locally deposited onto the solid surface or solid matter is eroded and added to the fluid the concentration field is changed locally by the mass difference $C(\vec{x}, t+1) = C(\vec{x}, t) +\delta m(\vec{x},t)$. For this mass in-/ejection not to introduce any artificial momentum into the fluid, the mass change is equally distributed according to the weights: $g_i(t+1) = g_i(t) + \omega_i \cdot \delta m$. This guarantees that the local momentum of the exchanged mass is zero as we have $\vec{p}_{\delta C} =  \delta m \sum\limits_i\omega_i \cdot \vec{c}_i = \vec{0}$.

\subsection{Calculation of the Wall Shear Stress}
\label{sec:calcwss}
The tangential wall shear stress acting on a surface is calculated from the shear stress tensor $\sigma$ given by the fluid. There are two ways to compute the shear stress tensor. For the LBM the shear stress tensor can be calculated from the non-equilibrium part of the distribution functions:
\begin{equation}
\label{eq:simpleshear}
\sigma^{ab}(\vec{x}) = \bigg ( 1-\frac{1}{2\tau} \bigg ) \cdot\sum\limits_i f^{neq}_i \cdot (\vec{c}_{i})^a(\vec{c}_{i})^b ,
\end{equation}
where $f^{neq}_i = f_i - f^{eq}_i $ and $(\vec{c}_{i})^a $ denotes the component $a$ of the discrete velocity vector $i$. Additionally, Thampi et al. \cite{Thampi20131} propose a very accurate way to calculate derivatives of physical properties in LBM. From this we can calculate the shear tensor as:
\begin{equation}
\label{eq:sauronshear}
\sigma^{ab} = \frac{\nu}{T} \sum\limits_i \omega_i \rho(\vec{x}+\vec{c}_i) 
\left(u^a(\vec{x}+\vec{c}_i)(\vec{c}_i)^b + u^b(\vec{x}+\vec{c}_i)(\vec{c}_i)^a \right) ,
\end{equation}
where the prefactor $T$ depends only on the weights and discrete velocity vectors and needs to be computed just once:
$$
T = c_s^2 \sum\limits_i \omega_i \vec{c}_i \cdot \vec{c}_i .
$$
We found that Eq. \eqref{eq:sauronshear} is more accurate for calculating the shear tensor, but it requires that all neighboring nodes are fluid nodes. Therefore we use Eq. \eqref{eq:sauronshear} wherever possible and Eq. \eqref{eq:simpleshear} elsewhere. The wall shear force can be calculated as the product of the normal vector $\vec{n}$ on the boundary surface times the deviatoric shear stress $\vec{\tau} = \vec{n}\cdot \sigma$. Since the boundary in general does not lie on a fluid node we linearly extrapolate the shear stress onto the boundary as follows:
$$
	\vec{\tau}_w = \frac{\sum_i^{N_f} \left[ (1 + \Delta) \vec{\tau}(\vec{x})  - \Delta \vec{\tau}(\vec{x}+\vec{c}_i) \right]}
	{N_f},
$$
where $\vec{\tau}(\vec{x})$ is the shear force at the interface cell, $\Delta = 1-m(\vec{x})$ is the relative distance from the node to the wall and the sum runs over all neighbors ($\vec{x}+\vec{c}_i$) that are fluid nodes.
For the erosion and deposition law we only take into account the absolute value of the tangential component of the wall shear force (see Eq. \eqref{eq:tang}). The unitary normal vector to the surface can be calculated by the color gradient $\hat{n}$ of the mass field (see appendix  \ref{app:colorgradient}). The normalized surface vector must be multiplied by the surface area to find the total WSS acting within an interface cell onto the surface, the total surface vector is then $\vec{n} = \hat{n}\cdot\delta\Omega$.
For each interface node we need to estimate the surface area that is exposed towards the fluid, this area $\delta\Omega$ is estimated as the area of the plane defined by the surface vector $\hat{n}$, cut by the cells boundary, hence the surface is piecewise flat. Therefore the area is unity if the surface is parallel to the lattice planes and is maximum $\sqrt{3}$, e.g. when $\hat{n} = (1,1,1)/\sqrt{3}$.  The total mass exchange caused by the erosion and deposition law within an interface cell is then $\delta m(t) = \dot{m}(t)\cdot \delta\Omega \cdot \delta t$.

\subsection{Boundary Interpolation Scheme}
\label{sec:interp}
\begin{figure}
\includegraphics[width=\columnwidth,trim = 1cm 0 0 0]{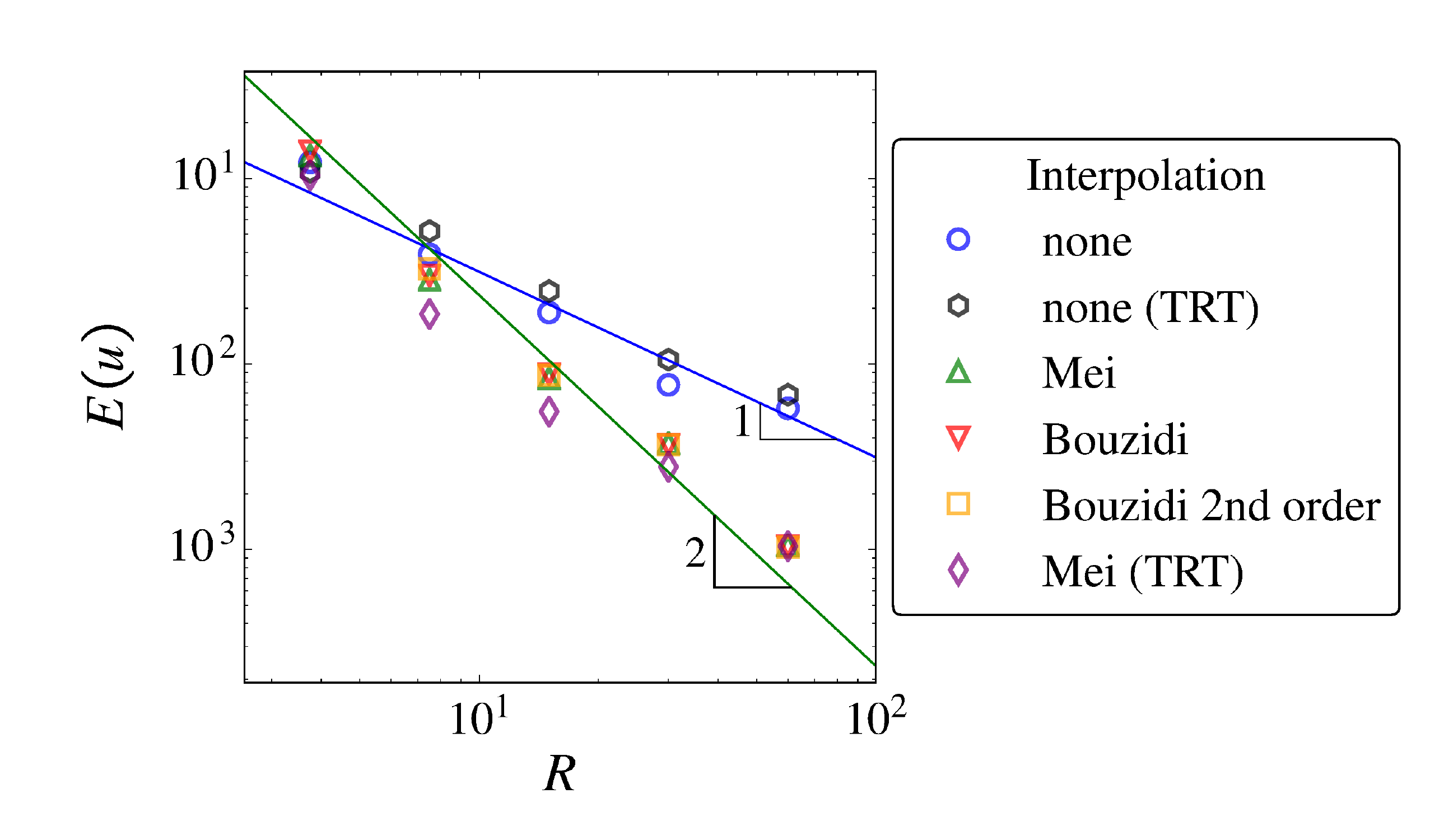} 
\caption{Poiseuille flow through a pipe with a given driving force $10^{-6}$ and $\mathcal{T}=0.6$ (in lattice units). The two-norm of the error $E(u)$ (see appendix \ref{app:twoerr}) of the flow field for the Bouzidi and the Mei interpolation schemes and the half-way bounce back, combined with the BGK or the TRT (denoted by TRT) collision operator, are compared. The interpolation schemes show similar accuracy and scale better than the simple bounce back.}
\label{fig:uerr}
\end{figure}

The commonly used no-slip boundary condition in LBM is imposed by the bounce-back method, which puts the boundary either on lattice nodes (on-site bounce-back) or in the middle of two lattice nodes (half-way bounce-back). In order to allow for boundaries that are more generic and do not lie on points restricted by the lattice one has to modify the bounce-back method. There are several interpolation schemes to account for generic boundaries. Two of the most widely used schemes are from Bouzidi et al. \cite{bouzidi} and Filippova et al. \cite{Filippova1998219}. The interpolation scheme from Filippova was improved by Mei et al. \cite{Mei2000680} to allow for lower viscosity. We implemented both methods and compared them simulating the Poiseuille flow through a pipe of radius $R$. The analytical solution for the velocity profile of the Poiseuille flow is as follows:
$$
u(r) = \frac{\Delta P }{4\rho\nu L}(R^2 - r^2),
$$
where $r$ is the distance to the center of the pipe, $R$ is the radius of the pipe, with a fluid driven by a pressure drop $\Delta P$ over length $L$. For a force driven fluid the pressure drop can be substituted by the driving force, $\Delta P/L \to F$. For simulating the Poiseuille flow, we set the mass field to zero inside $m(r<R) = 0$ and outside to one $m(r>R) =1$. For interface cells the mass depends on the distance to the center $m(r\simeq R) = 1-(R-r)$. Here we switch off the erosion and deposition of solid material. The fluid is driven by a constant force in the direction of the pipe and we employ periodic boundary conditions at the inlet and outlet, hence only a one layer cross-section of the pipe has to be stored. The two-norm of the error of the stationary flow field $\vec{u}$, as defined in appendix \ref{app:twoerr}, was compared for the two interpolation schemes and the half-way bounce back (see figure \ref{fig:uerr}). It was found that both interpolation schemes show very similar accuracy (second order) and are indeed superior to the classical half-way bounce back scheme (first order).

These interpolation schemes can also be used together with the TRT collision operator and we observe that there is no considerable difference in accuracy when compared with BGK (see figure \ref{fig:uerr}). For stability reasons we settled on using the interpolation by Mei for our model.

\subsection{Validation}
\label{sec:val}
To verify that our model for erosion and deposition is working properly a setup is used where the flow field, the WSS, and hence also the surface evolution can be calculated analytically. For this purpose we consider pipe flow where the fluid is described by Poiseuille flow. The setup used for the simulations is the same as already described in section \ref{sec:interp}, except that now the interface cells and hence also the surface evolve according to the erosion and deposition laws as described in section \ref{sec:erdep}. 

\subsubsection{Validating Erosion: Erosion of a Pipe}
\label{sec:vererosion}
\begin{figure}
\includegraphics[width=\columnwidth]{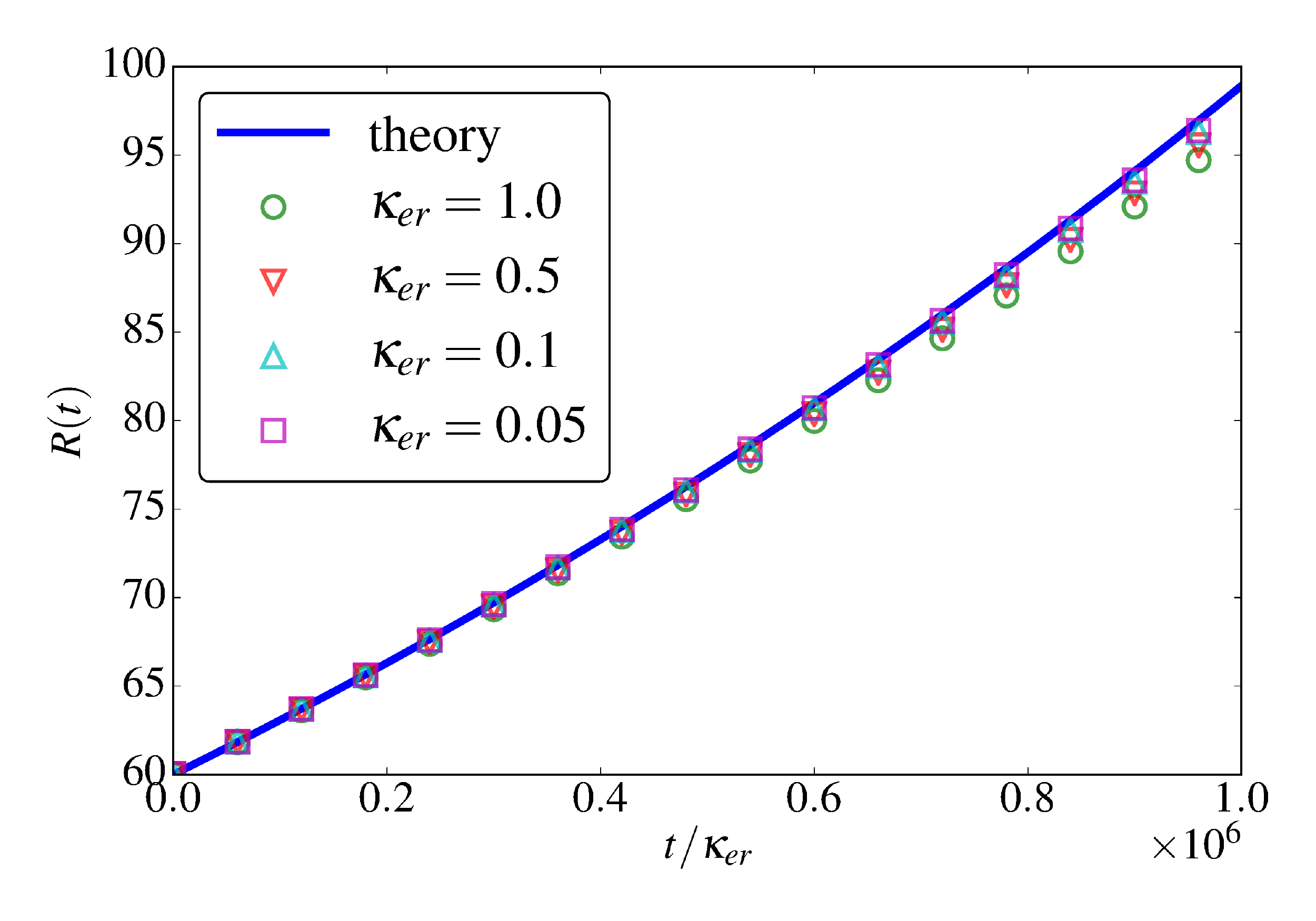}
\caption{Pipe erosion for different erosion coefficients. Initial pipe radius is $R(0)=60$ (in lattice units), the erosion was started after a steady flux $\Phi = \int_\Omega \rho u$ was reached ($ (\Phi (t+1)-\Phi(t) )/ \Phi (t) < 10^{-10} $). The time is rescaled by the erosion coefficient $\kappa_{er}$. A relaxation time $\mathcal{T} = 0.6$ was used, the erosion threshold is zero, $\tau_{er}  = 0$, and the driving force is $10^{-6}$ (in numerical units). The theoretical curve (blue line) is given by equation \eqref{eq:pipegrow}. }
\label{fig:pipeerosion}
\end{figure}

\begin{figure}
\includegraphics[width=\columnwidth]{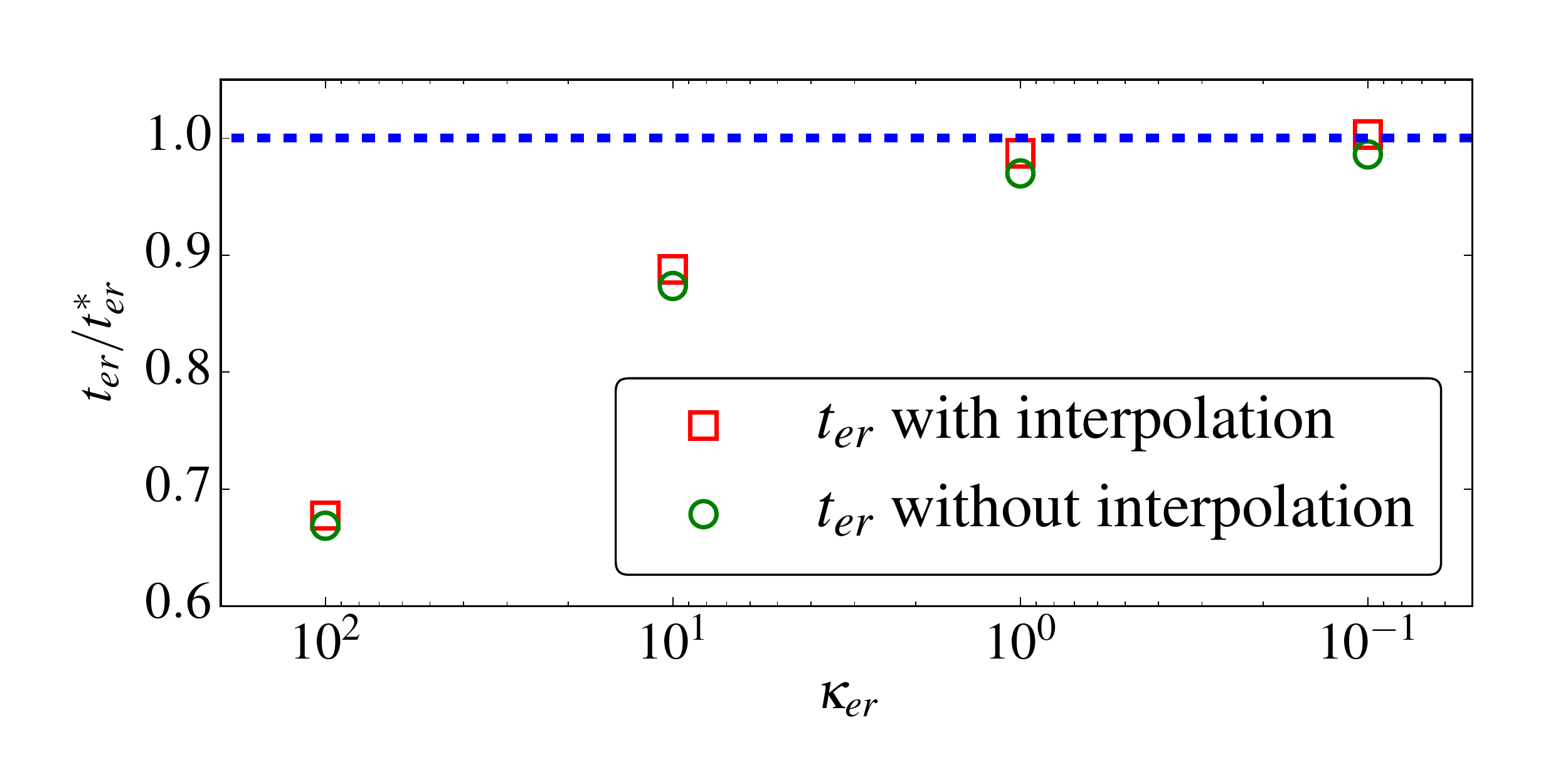}
\caption{Convergence of the erosion time for small erosion coefficients. The plot shows the difference between the measured erosion time and the theoretical one $t_{er}/t_{er}^* $. The setup is the same as in the simulations shown in figure \ref{fig:pipeerosion}, only the erosion coefficients are different. We see that the boundary interpolation scheme improves the accuracy of the erosion time $t_{er}$. The theoretical erosion time is $t^*_{er} = \rho_g/\kappa_{er}|\vec{F}|$, denoted by the dashed line. The erosion time $t_{er}$ is measured by fitting the measured pipe radii (see figure \ref{fig:pipeerosion}) to the analytic equation \eqref{eq:pipegrow}.
} 
\label{fig:ercoeff}
\end{figure}

To measure the erodibility of cohesive soils the hole erosion is commonly used (see e.g. Ref. \cite{wahl2010comparison}). In this test water is flowed through a soil pipe under constant pressure drop and the soil is eroded and thus the pipe grows in width. Bonelli et al. \cite{Bonelli2006555} developed an analytical theory to predict the evolution of the pipe given specific properties of the soil and fluid. They found a scaling law for the evolution of the radius that depends on the pressure $\Delta P$ driving the fluid, the critical shear stress $\tau_c$ and the erosion coefficient $\kappa_{er}$:
\begin{equation}
\label{eq:pipegrow}
	\frac{R(t)}{R_0} = 1 + \left(1 - \frac{\tau_c}{\Delta P} \right) \left[\exp\left(\frac{t}{t_{er}}\right) -1 \right], 
\end{equation}
the erosion time is $t_{er}= \rho_gL/\Delta P \kappa_{er}$ with the density $\rho_g$ of the erodible soil.
We use the same setup except that instead of a pressure difference driving the fluid we introduce a constant driving force $|\vec{F}| \equiv \Delta P / L $ in the direction of the pipe. The force is implemented for the BGK as derived by Guo et al. \cite{PhysRevE.65.046308} and for the TRT as described by Seta et al. \cite{seta2014implicit}. The pipe erosion law derived by Bonelli shows good compliance with hole erosion experiments and is analytically exact for slow erosion. The wall shear stress depends only on the driving force and pipe radius: $\tau_w = |\vec{F}| \cdot R(t) /2$. For low erosion coefficients our simulations show excellent agreement with the erosion scaling law as can be seen in figure \ref{fig:pipeerosion}. The flow is quasi-steady and for all times close to the analytical Poiseuille profile. However as the erosion rate increases with larger pipe radius the condition of slow erosion breaks down at some point. We can observe in figure \ref{fig:ercoeff} that for low erosion coefficients ($\kappa_{er} \to 0$) the simulated erosion of the pipe is close to the theoretical scaling but deviates for higher erosion coefficients.

\subsubsection{Validating Deposition: Clogging of a Pipe}
\label{sec:verdep}
To verify that our model also works properly for deposition, we start with an initial steady Poiseuille flow through a pipe, set a constant particle concentration and set the deposition threshold above the wall shear stress. For Poiseuille flow we know that the WSS is $\tau_w = |\vec{F}| \cdot R(t)/2$ and the deposition law is given by $\dot{m} = (\tau_{dep}-\tau_w)\cdot C \cdot \kappa_{dep}$. Thus we can derive a differential equation for the radius:
$$
	\dot{R}(t) = \kappa_{dep}C \bigg( \frac{|\vec{F}| R(t)}{2}- \tau_{dep} \bigg) ,
$$
which is solved by the exponential function:
\begin{equation}
\label{eq:rdep}
	R(t) = \left(R_0 - \frac{2\tau_c}{|\vec{F}|} \right) \exp\bigg({\frac{|\vec{F}|\kappa_{dep}C}{2}t}\bigg) + \frac{2\tau_{dep}}{|\vec{F}|} ,
\end{equation}
where we assume a constant concentration $C$ over time.

\begin{figure}
\includegraphics[width=\columnwidth]{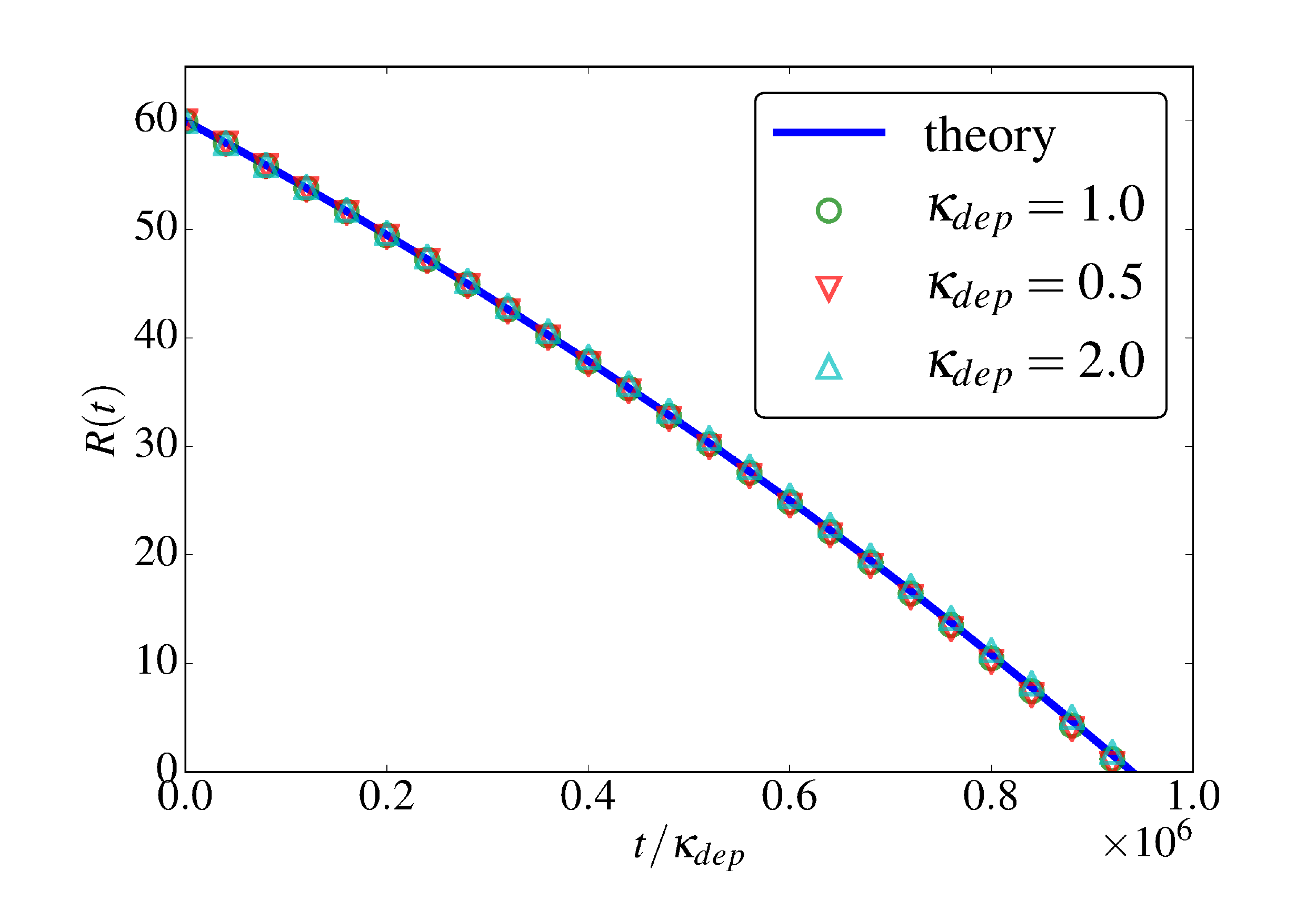}
\caption{A pipe with initial radius 60 (lattice units) is clogged by the matter suspended within the fluid, the deposition threshold is set to $\tau_{dep}=8\cdot 10^{-5}$, the relaxation time is $\mathcal{T}=0.6$, the driving force $10^{-6}$ (in numerical units) and the concentration to $C=1$. The smaller the radius becomes, the lower is the WSS and the deposition will attain a linear behavior $\dot{m} \to (\tau_{dep})\cdot C \cdot \kappa_{dep}$. The theoretical expression is given by equation \ref{eq:rdep}.
}
\label{fig:pipeclog}
\end{figure} 

In figure \ref{fig:pipeclog} we see that the simulations are in good agreement with the analytic prediction, however this prediction is only valid for slow deposition where the flow field is quasi-steady and therefore close to the Poisseuille profile at all times.

\section{Channelization in Porous Media}
\label{sec:results}
Our model is now applied to simulate erosion and deposition in a soft porous medium, by soft we mean that all solid matter within the simulation is erodible. For these simulations we use two basic setups, the first is to impose a constant pressure drop between the inlet and outlet, and the second enforces a constant flux at the inlet. For both setups the fluid driving forces are assumed to be much stronger than gravitational forces and therefore the latter is neglected. The flux is imposed using inlet boundary conditions as described by Kuzmin et al. \cite{Kuzmin2219}. Due to computational power restrictions it is challenging to simulate a porous medium of the same size as occurring in nature or in experiments. To minimize the finite size effects we employ periodic boundary conditions at the walls. The porous medium consists of randomly placed spheres (of radius 7.1 grid points), if a sphere is cut by a wall, it will continue on the opposite side of the wall. Effectively this setup is equivalent to a porous medium with recurring pattern and infinite width. Spheres can also overlap and are placed until a porosity of 0.5 is reached. The inlet and outlet regions ($z < 20$ and $z > 120$ cells, respectively) are kept free of solid matter. The dimensions of the simulation box are $100 \times 100 \times 140$ ($xyz$, where $z$ is the flow direction). The particle concentration is fixed to $C=0.1$ at the inlet, however it changes within the porous medium due to deposition and erosion. Since we are interested in laminar flow; and higher relaxation time parameters achieve faster convergence to stationary states; the relaxation times are set to $\mathcal{T}=\mathcal{T}_s=1$ for all following simulations. For both setups we first use a sharp transition between deposition and erosion, i.e. $\tau_{er}=\tau_{dep}$. 

\subsubsection{Constant Pressure Drop}
\label{sec:drop}
\begin{figure}
\includegraphics[width=\columnwidth]{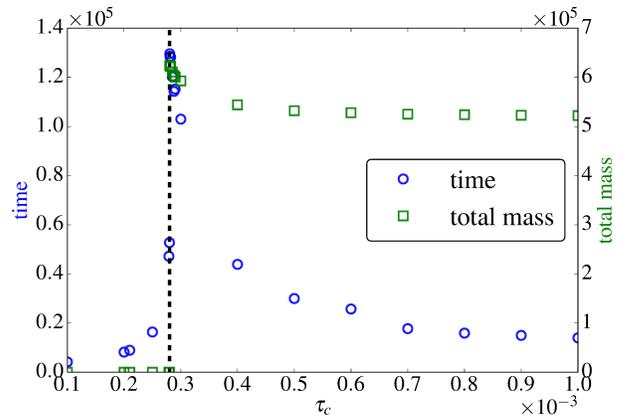}
\caption{A porous medium with initial porosity $0.5$ is evolved by a fluid that is driven by a relative pressure drop of $0.02$. The inlet solute concentration is $0.1$, the coefficients $\kappa_{dep}=10^2$, $\kappa_{er}=10$ and $\mathcal{T}=\mathcal{T}_s =1.0$. We measure the time until there is no more deposition or erosion and the total mass of erodible matter contained in the final structure. The critical threshold (dashed vertical line) is at $2.803 \pm 0.002 \cdot 10^{-4}$. Below the critical threshold, the medium ends up completely eroded, there is no more solid mass in the simulation; above, it ends up clogged. The size of the simulation box is $100\times100 \times 140$, where the inlet and outlet (each a length of $20$ cells) are constrained from deposition to avoid changing the inlet boundary conditions of the fluid. While specific numbers for time and total mass change with the system size, the two regimes of clogging and complete erosion remain.
}
\label{fig:phead}
\end{figure}

\begin{figure}
\includegraphics[width=\columnwidth]{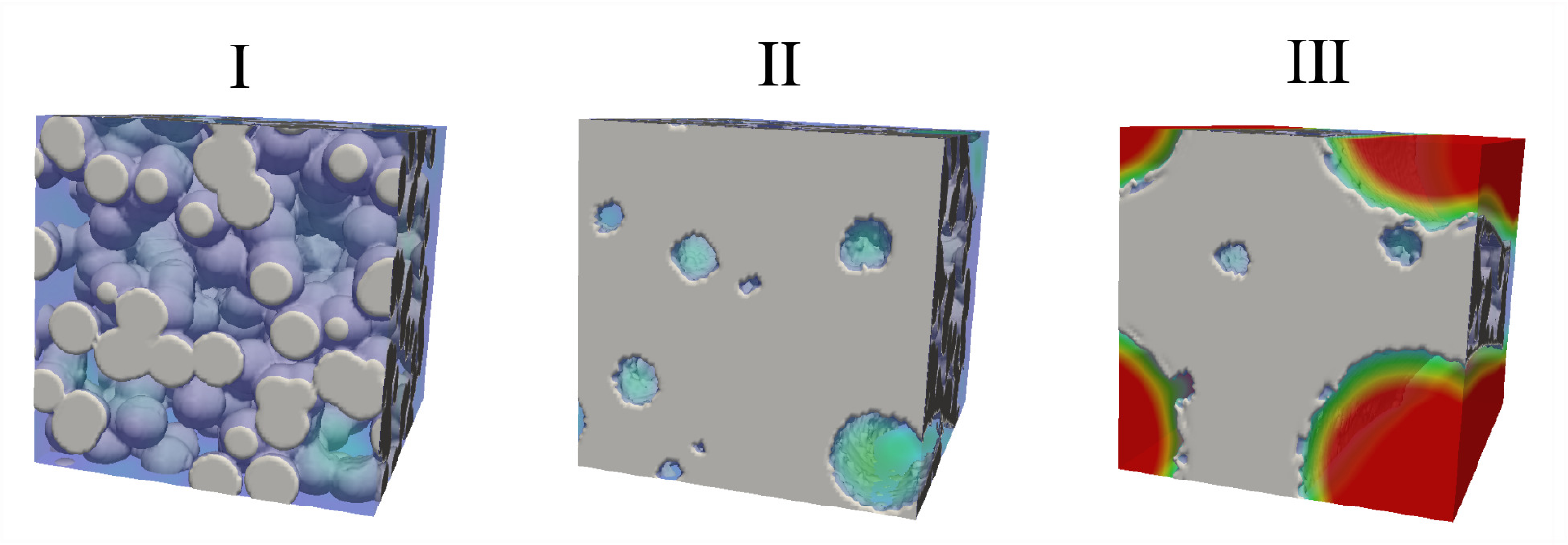}
\caption{Snapshots of the evolution of the porous medium for constant pressure drop and the shear threshold $\tau_c=2.8\cdot 10^{-4}$ just below the critical value (see figure \ref{fig:phead}). The solid domain is shown in solid gray, the fluid in transparent color; low velocity fluid is shown in light blue, medium velocity in green, and high velocity in red. In this case the medium ends up completely eroded. Snapshot I ($t=1$) shows the initial stage, II ($t=3 \cdot 10^5$) shows when discrete channels are formed, and at stage III ($t=4.5 \cdot 10^5$) we see that smaller channels vanish while the biggest one grows.
}
\label{fig:pheadpics}
\end{figure}

For a constant pressure drop we find that there is a critical value for the erosion/deposition threshold (see figure \ref{fig:phead}). When the threshold is set below this critical value, the porous medium will completely erode and the flux will diverge. On the other hand, if it is set above the critical value the porous medium clogs completely and the flux goes to zero. We also see that the total time, until a final state is reached, increases when we go from eroded to clogged final states (see figure \ref{fig:phead}). The reason behind this is simple, clogging will reduce the flow and slow down the process, whereas erosion increases the flux and thus is a diverging process. The total mass of a clogged medium increases for low thresholds, but only slightly (see figure \ref{fig:phead}), this is due to a slower clogging process, which allows to fill up more pores with solid matter. The time it takes for the system to reach its final state grows the closer the threshold is to the critical one. However, before the porous medium is completely eroded or clogged, we see the formation of channels within the porous structure (see figure  \ref{fig:pheadpics}). The size of the largest channel formed will decide whether the medium will end up eroded or clogged, as already seen in the verification examples (section \ref{sec:val}), a pipe and analogously also a channel with a constant pressure drop will always continue to clog or erode (see appendix \ref{app:pdrop}). The critical threshold is however not generic, it depends not only on external parameters but also on the initial porosity and on the specific inner structure of the porous medium. For example, it increases for higher pressure drop or larger pores, because the occurring shear also increases. As will be seen in the next application, there is in theory a critical pipe radius for which there is neither erosion nor deposition, however such a configuration is very unstable and even small fluctuations in the flux or shape will lead to an outcome of complete erosion or clogging (see appendix \ref{app:pdrop}). Therefore we can conclude that when considering a sharp transition between erosion and deposition, there are no configurations for constant pressure drops that lead to steady non-trivial structures.

\subsubsection{Constant Flux}
\label{sec:cflux}
\begin{figure}
\includegraphics[width=0.5\columnwidth]{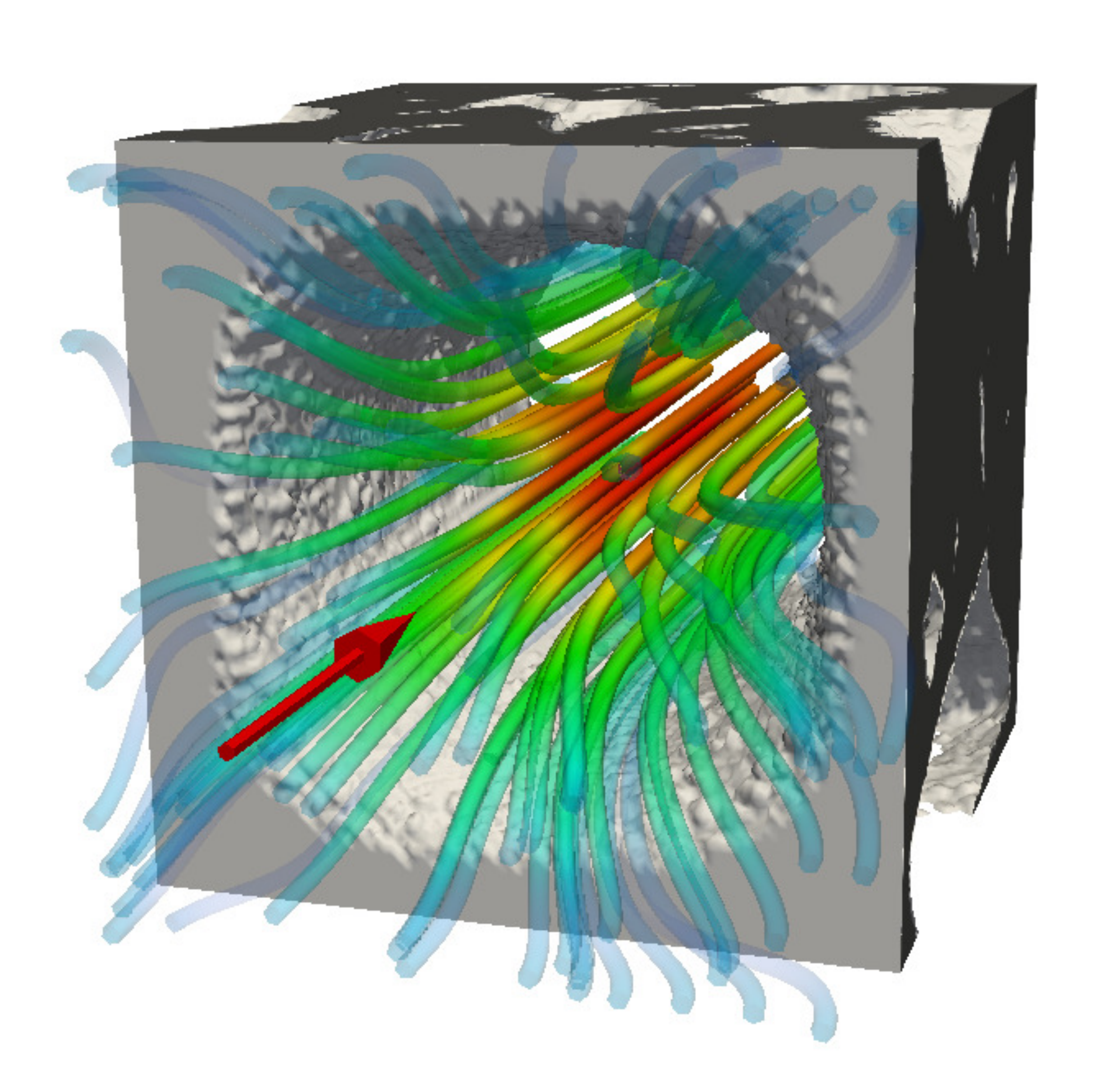}
\caption{Steady one channel state after erosion and deposition for a threshold $\tau_c = 5\cdot 10^{-5}$. The flow direction is annotated by the (red) arrow. At the inlet the shape is more squared and wider, further inside the channel narrows and its shape more closely resembles a circular pipe. 
}
\label{fig:onechannel}
\end{figure}

Contrary to the constant pressure drop, a constant flux leads to a final state of the porous medium with stable channels. For the case of a sharp transition from deposition to erosion ($\tau_{dep}=\tau_{er}$) the medium ends up having only one channel. As can be expected due to symmetry reasons it closely resembles a pipe, as the pipe is the only channel form that has a uniform constant WSS for steady flow. The flow then also closely resembles Poiseuille flow, the difference to a perfect pipe stems from the fact that the inlet and outlet are not a pipe but rectangular by construction and so disturb the form slightly, see figure \ref{fig:onechannel} as an example.
We have found that the pipe is a bit wider at the inlet and then narrows slightly towards the outlet. 

\begin{figure}[ht!]
\includegraphics[width=\columnwidth]{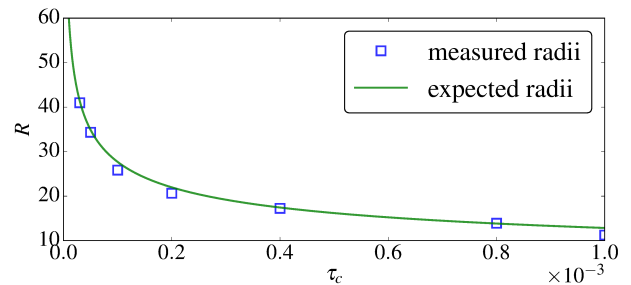}
\caption{At the inlet a constant flux $\Phi=10$ with constant velocity $u=0.001$ is imposed, the inlet solute concentration is set to 0.1 and the relaxation parameters to one, $\mathcal{T} = \mathcal{T}_s = 1$. The erosion and deposition coefficients are the same as in the previous setup (fig. \ref{fig:phead}). The radius of the channel is measured in the middle of the simulation box ($z=70$), the expected radii are calculated from equation \ref{eq:critrad}. The system has to be larger than the diameter of the pipe, otherwise additional boundary effects will occur.
}
\label{fig:cflux}
\end{figure}

The pipe radius was measured in the middle of the simulation box ($z=70$), which is of the same size as for the constant pressure drop simulations ($100\times100\times140$). For Poiseuille flow we know that the relation between radius and flux is given by $\Phi=(\pi/8\mu)(\Delta P/L)\cdot R^4$ and the WSS by $\tau = (\Delta P/L)(R/2)$, from which we can find the critical radius 
\begin{equation}
\label{eq:critrad}
R_c=\left(\frac{4\mu\Phi}{\pi\tau_c}\right)^{\frac{1}{3}} \quad .
\end{equation}
The critical WSS $\tau_c$ is the same as the erosion and deposition threshold $\tau_c=\tau_{er}=\tau_{dep}$. The formula for the critical radius shows that the radius depends on the imposed flux $\Phi$, the dynamic viscosity $\mu$ ($=\rho\nu)$ and the shear threshold $\tau_c$ and we can show that a pipe with this radius is indeed stable (see appendix \ref{app:stability}). We compare the measured radius of the channel with the expected radius and find that they are in very good agreement (see figure \ref{fig:cflux}). Theoretically, having two or more pipes with the same radius could also be a steady configuration, it would however only be meta-stable, small fluctuations would result in clogging one pipe and widening the other (see appendix \ref{app:stabilitytwo}). 

\subsubsection{Constant Flux and Gap}
\begin{figure}[ht!]
\includegraphics[width=\columnwidth]{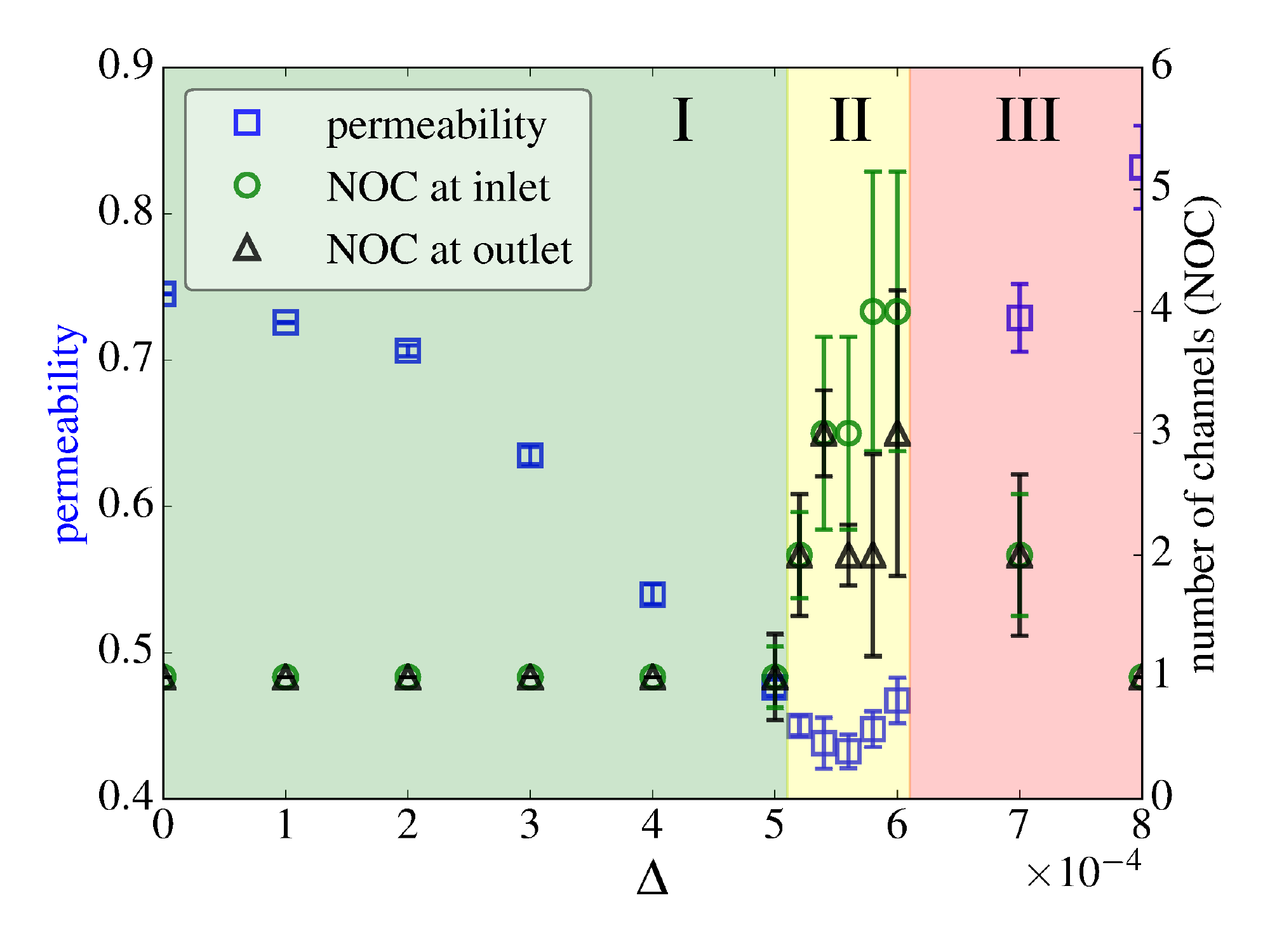}
\caption{Starting from the simulation shown in figure \ref{fig:cflux} with the threshold $\tau_c=4 \cdot 10^{-4}$ we introduce and increase a gap $\Delta$ between the deposition $\tau_{dep}=\tau_c-\frac{\Delta}{2}$ and the erosion threshold $\tau_{er}= \tau_c+\frac{\Delta}{2}$. In addition to measuring the final permeability, the number of channels was counted at the inlet and the outlet. For each gap, we made simulations for four random porous media made of solid spheres (with radius 7.1 grid points) with different seeds for the random number generator (for the placement of the spheres). The first range (I) includes final states with only one channel, in the second range (II) there are multiple channels and bifurcations. In the third range (III) channels are not distinguishable anymore, but the medium is a porous structure that depends strongly on the original structure.
}
\label{fig:cfluxgap}
\end{figure}

After finding the steady state solutions for a sharp transition we now study a system with a gap between deposition and erosion. This means that there is a range of WSS for which matter is neither removed nor attached. For small gaps we found that the final configuration still features one channel (see figure \ref{fig:cfluxgap}), which however deviates more and more from a perfect pipe, the wider the gap, its form depending on the original porous structure. For even larger gaps we found that there are steady states with more than one channel, the number of channels depends not only on the gap but also on the flux and the size of the simulation box. The smaller the pipe formed by erosion/deposition without a gap, the more channels can be expected for a final state with gap. We can also see in figure \ref{fig:cfluxgap} that the permeability monotonously decreases for increasing gap until there is a minimum, which is also where many distinct channels are formed. The permeability increases again when going to even larger gaps which leads to final structures that do not show distinguishable channels but depend strongly on the original porous structure. For very large gaps there is neither erosion nor deposition and the porous structure is static, in the setup of figure \ref{fig:cfluxgap} this is the case for gaps larger than $8\cdot 10^{-3}$. Note that $\Delta=0$ corresponds to the solution shown in figure \ref{fig:cflux}, while changing $\tau_c$ or the system size changes the permeability and the number of channels; it also limits the range of $\Delta$; the qualitative behavior is the same as long as $\tau_c$ is such that both erosion and deposition are active. The final state depends strongly on the local deposition and erosion, the stronger the flow changes the porous structure, the less the final state depends on the original structure. Although the gap in this case cannot be adjusted, the final structure can be influenced by adjusting the flow velocity and hence the shear within the porous medium. Mahadevan et al. \cite{mahadevan2012flow} concluded that channelization can arise due to the preferential erosion induced by the flow. As channelization cannot progress further when there is only one channel left we showed that this effect is stronger the smaller the difference between deposition and erosion thresholds. 

\begin{figure}[ht!]
\includegraphics[width=\columnwidth]{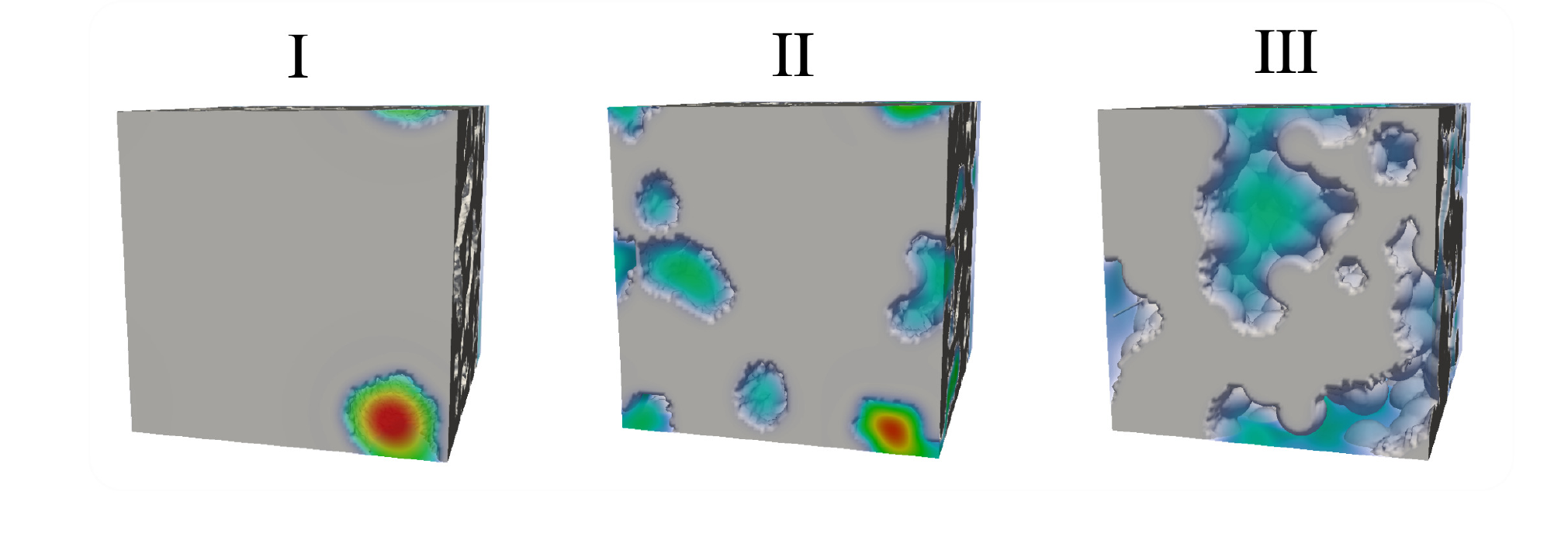}
\caption{Snapshots of the final state of the solid structure after erosion and deposition for different gaps $\Delta$ between erosion and deposition thresholds. These snapshots show examples for the three different regions painted in figure \ref{fig:cfluxgap}. The first snapshot (I) corresponds to a gap of $\Delta = 4\cdot 10^{-4}$, the second (II) to $\Delta = 5.4\cdot 10^{-4}$ and the third (III) to $\Delta = 7\cdot 10^{-4}$.
}
\label{fig:gapshots}
\end{figure}

\section{Conclusion}\label{conclusions}
We have found that erosion and deposition within porous media always lead to channelization. However stable channels for any material can only be expected if the fluid features a constant flux. For a fluid driven by constant pressure drop, the final configuration is either completely clogged or completely eroded. On the other hand a constant flux will always leave at least one stable channel. This statement is also supported by the experimental findings of Alem et al. \cite{Alem2015}, they tested deposition in a filter, and found only completely clogged final states for constant pressure drop, but steady state configurations for constant flux. We also showed that the final state of a porous medium altered by erosion and deposition crucially depends on the balance between the two effects and a circular pipe is formed when the transition is sharp. While specific values such as permeability or number of channels change with the system size; and the threshold for erosion and deposition; the qualitative behavior remains the same. Our numerical findings may help to better understand and mitigate internal erosion in embankment dams \cite{erosiondams} and also in naturally occurring flows through porous media. 
In our model gravity is not considered, if gravitational forces are not negligible, buoyancy and sedimentation of suspended particles play a role, hydrostatic pressure increases towards the lower part of the fluid and also the rate of deposition changes \cite{PhysRevLett.97.138001}. To study these effects in the future, it is possible to incorporate gravitational forces into our model by adding an external force to the fluid dynamics equations and by making the deposition dependent on particle density.

While our model allows to simulate erosion and deposition for a variety of fluids and materials, it cannot capture other interesting phenomena in flows through porous media such as decompaction and fluidization \cite{PhysRevE.78.051302}. However, extensions of the model to study these effects can be a subject of future research. 

\begin{acknowledgments}
We acknowledge financial support from the European Research Council (ERC) Advanced Grant 319968-FlowCCS.
\end{acknowledgments}

\appendix
\section{Appendix}\label{sec:app}

\subsection{Color Gradient}
\label{app:colorgradient}
The solid and fluid domains are defined by the mass scalar field $m(\vec{x})$ (see section \ref{sec:mass}). To get the normal vector of the surface we define the color gradient in the following way:
$$
	\hat{n}(\vec{x}) := -\frac{\sum\limits_{i} m(\vec{x}+\vec{c}_i)\cdot \vec{c}_i}
						{\sum\limits_{i} m(\vec{x}+\vec{c}_i)\cdot ||\vec{c}_i||_2}
$$
This equation defines the orientation of the surface and allows to compute the normal and tangential part of fluid properties, e.g. the shear force.

\subsection{Error measurement}
\label{app:twoerr}
To measure the difference between an analytic velocity field $\vec{u}_{theory}$ and a numerical computed field $\vec{u}_{sim}$ a measure for the error must be defined. It is common practice to use the two-norm of the difference:
$$
E(u) := \sqrt{\frac{\sum\limits_{\vec{x}}\big\{\vec{u}_{sim}(\vec{x})-\vec{u}_{theory}(\vec{x}) \big\}^2}
{\sum\limits_{\vec{x}}\vec{u}^2_{theory}(\vec{x})}},
$$
where the sum runs over all nodes in the fluid domain. This measure allows to evaluate the accuracy of a fluid dynamics computation (see figure \ref{fig:uerr}).

\subsection{Constant pressure drop channel instability}
\label{app:pdrop}
For Poiseuille flow through a pipe the wall shear stress is given by:
$$
\tau = \frac{\Delta P}{L} \frac{R}{2}
$$
The critical radius for which there is neither erosion nor deposition would be $R_c =  2 \tau_c L/(\Delta P)$. However, for a sharp transition between erosion and deposition ($\tau_c=\tau_{er}=\tau_{dep}$) a small fluctuation in radius or pressure difference will lead to complete erosion or complete clogging:
\begin{align}
	R > R_c &\Rightarrow \tau(R) > \tau_c &\Rightarrow \dot{m} < 0 &\Rightarrow \dot{R} > 0 \\
	R < R_c &\Rightarrow \tau(R) < \tau_c &\Rightarrow \dot{m} > 0 &\Rightarrow \dot{R} < 0
\end{align}
These equations illustrate why a channel is unstable under constant pressure drop (see section \ref{sec:drop}). \\

\subsection{Constant flux channel stability}
\label{app:stability}

For a Poiseuille flow through a pipe with constant flux, the flux is determined by $\Phi=(\pi/8\mu)(\Delta P/L) R^4 $, and the shear by $\tau = (\Delta P/L) (R/2)$. Thus, we can derive an equation for the shear depending on the flux:
$$
\tau(R)=\frac{4 \Phi \mu}{\pi R^3}.
$$
From this it is easy to see that the pipe with critical radius $R_c$ is stable, for a sharp transition from erosion to deposition ($\tau_c=\tau_{er}=\tau_{dep}$) :

\begin{align}
	R > R_c &\Rightarrow \tau(R) < \tau_c &\Rightarrow \dot{m} > 0 &\Rightarrow \dot{R} < 0 \\
	R < R_c &\Rightarrow \tau(R) > \tau_c &\Rightarrow \dot{m} < 0 &\Rightarrow \dot{R} > 0
\end{align}

These equations show that for the same flux a larger pipe will result in lower flow velocity and thus in a lower shear, which results in deposition; and vice versa for pipes with radii smaller than the critical one. This is the reason why a constant  flux will always form one stable channel (see figure \ref{fig:cflux}).

\subsection{Stability analysis for a two channel state}
\label{app:stabilitytwo}
A final state with two (or more) pipes is theoretically possible for sharp transitions ($\tau_c = \tau_{er} = \tau_{dep}$) and would yield a critical radius of 
$$
R_{c,2} = \bigg(\frac{2\mu\Phi}{\tau_c\pi}\bigg)^\frac{1}{3} ,
$$
where we assume the pipes lie far apart and hence assume a Poiseuille profile for both flows. This is stable for perturbations that are symmetrically occurring for both pipes. However a perturbation of just one pipe will widen one and clog the other pipe. Assume there is a perturbation of flux and the flux in one channel increases by $\delta \Phi$, since the total flux is constant, the flux in the other channel decreases by the same amount. The WSS in the first channel now increases:
$$
	\tau_1 \to \frac{2\mu (\Phi+\delta \Phi)}{\pi R_{c,2}} > \tau_c,
$$
thus the first channel will start to erode and the second will clog until it vanishes. Even though this is only a sketch it shows that configurations with two (or more) channels are unstable.

\subsection{Conversion from numerical to physical units}
\label{app:units}
For comparing numerical with experimental findings, numerical units have to be converted to physical units, a very instructive recipe how to do this is written by J. Latt \cite{latt2008choice}. Here we provide an example for converting the units of the simulation shown in figure \ref{fig:phead}, for this purpose we match three parameters, namely the fluid density $\rho$, the viscosity $\nu$ and the length scale $\delta x$. The parameters are matched using the assumption that the diameter of the spheres building the porous medium is 1.4 mm; and the fluid is a heavy crude oil \cite{Hart2014} with density $\rho=1000 $ kg/m$^3$ and viscosity $\nu=10^{-3}$ m$^2$/s. Hence all other parameters can be calculated using the unit conversion scheme shown in \cite{latt2008choice}, e.g. for the time scale:
\begin{equation}
\label{eq:unitconv}
\nu_p = \nu_l \frac{\delta x ^2}{\delta t} \implies \delta t = \frac{\nu_l}{\nu_p} \delta x ^2 ,
\end{equation}
where $\nu_l$ is the viscosity in numerical units, and $\nu_p$ in physical units.

\begin{table}[h]
\begin{center}
  \begin{tabular}{ | c | c | c | }
    \hline
    parameter & numerical units & physical units \\ \hline
    $\rho$ & 1 & $1000$ kg/m$^3$ \\ \hline
    $\nu$ & 1/6 & 10$^{-3}$ m$^2$/s \\ \hline
    $\delta x$ & 1 & 0.1 mm \\ \hline
    $\delta t$ & 1 & $1.7 \cdot 10^{-6}$ s \\ \hline
    $\kappa_{er}$ & 1 & 0.017 s/m \\ \hline
    $\tau_c $ & 2.8 $\cdot 10^{-3}$ & 9689 Pa \\
    \hline
  \end{tabular}
\end{center}
	 \caption{This table shows an example conversion from numerical to physical units. The first three parameters ($\rho,\nu,\delta x$) were set to match the fluid properties of a heavy crude oil and a porous structure of pore size 1.4mm, the other parameters ($\delta t, \kappa_{er},\tau_c$) can be calculated from the first three as shown in Eq. \eqref{eq:unitconv}.}
\end{table}

Note that $\kappa_{er}$ and $\tau_c$ are difficult to match with experimental data, since they depend highly on the material and the formation of the porous medium. For instance, while the erosion coefficient is close to values reported by Bonelli et al. \cite{Bonelli2006555}, the erosion threshold is almost two orders of magnitude higher.

\bibliographystyle{ieeetr}
\bibliography{citations.bib}
\end{document}